\documentclass [logo,url,a4paper,10pt]{ETHpaper}
\usepackage{times}
\usepackage{natbib}
\usepackage{graphicx}
\usepackage{color,soul}
\usepackage{amssymb,amsmath}

\begin{document}


\title{Hierarchical Consensus Formation Reduces the Influence of Opinion Bias}


\author{Nicolas Perony, Ren\'e Pfitzner, Ingo Scholtes, Claudio J. Tessone, Frank Schweitzer}
\address{Chair of Systems Design\\
ETH Zurich\\
Kreuzplatz 5\\
8032 Zurich\\
Switzerland\\
Email: \{nperony, rpfitzner, ischoltes, tessonec, fschweitzer\}@ethz.ch
}

\reference{\emph{Proceedings of the 26th European Conference on Modelling and Simulation}, pp. 662-668}

\date{} 

\maketitle 


\abstract 
We study the role of hierarchical structures in a simple model of collective consensus formation based on the bounded confidence model with continuous individual opinions.
For the particular variation of this model considered in this paper, we assume that a bias towards an extreme opinion is introduced whenever two individuals interact and form a common decision. As a simple proxy for hierarchical social structures, we introduce a two-step decision making process in which in the second step groups of like-minded individuals are replaced by representatives once they have reached local consensus, and the representatives in turn form a collective decision in a downstream process.
We find that the introduction of such a hierarchical decision making structure can improve consensus formation, in the sense that the eventual collective opinion is closer to the true average of individual opinions than without it. In particular, we numerically study how the size of groups of like-minded individuals being represented by delegate individuals affects the impact of the bias on the final population-wide consensus. These results are of interest for the design of organisational policies and the optimisation of hierarchical structures in the context of group decision making.

\section{Introduction}
Among the problems which apply to the dynamics of social organisations, one of the oldest and best-studied is that of collective decision making \citep[see for example the early work of][]{black1948}. The importance and the complexity of this problem become especially clear when considering the many instances in which groups made up of individuals with diverse opinions have to reach a common agreement, or {\em consensus}, on a particular question. Depending on the distribution of initial opinions and the propensity of individuals to adapt their opinion to that of others, consensus may eventually be reached or opinions within the group may polarise around a restricted number of distinct values, in which case the opinion space is said to be fragmented. The question of {\em if}, {\em when} and {\em where} in the opinion space a consensual agreement may emerge involves the study of the complicated interactions through which members of the group adapt their individual opinions. In other terms, an understanding is needed of the mechanisms by which social interactions between individuals may facilitate collective decision making processes. As argued for instance by \citet{Bonabeau2009}, one interesting aspect of such mechanisms is the fact that, under certain circumstances, they can alleviate the cognitive biases of individuals and thus result in better collective decisions. Naturally, depending on the type of individual biases present in the system, \emph{how} consensus is formed strongly affects how pronounced this beneficial effect is.
The goal of this paper is to study this question in the context of \emph{hierarchical approaches to consensus formation}, i.e. like-minded individuals gathering and forming a \emph{local} consensus. A single collective decision is then formed in a downstream process.

In order to address this question, we study how a simple proxy for a hierarchical decision process affects the influence of a systemic bias on the collective decision.
Our work is applied specifically to a modelling context, based on the well-known bounded-confidence model, which has been widely used as a standard model of opinion dynamics with constrained interactions between agents \citep{Krause2000,DefNeaAmbWei2000,dittmer2001,HegKra2002,WeiDefAmbNad2002,Weisbuch2004,Lorenz2007}. This model posits a population of individuals interacting within a continuous opinion space. In a variation of this model, we introduce a systemic bias which uniformly affects individuals whenever they interact with each other.
In the context of social systems, intuitive interpretations of this bias include for example the influence of predominant and highly-biased mass media, or the presence in a decision board of a strongly-opinionated member affecting the debate and inconspicuously steering the discussions. In our model, this influence increases the change in opinions of individuals when interacting with peers whose opinion is consistent with the bias. At the same time, it proportionally decreases the change in opinion of those individuals who interact with peers whose opinion is opposed to the bias. In the absence of any interactions, we assume that individual opinions are not affected by the systemic bias, thus resulting in a preservation of their \emph{status quo} as long as no interactions take place.

In this particular setting, we study how a two-step decision process based on the bounded confidence model impacts the eventual consensus reached within the population.
For this, we assume that groups of individuals -- after they have come to a group-wise, semi-collective decision in a first phase of \emph{bounded confidence} -- are represented by an aggregate, {\em representative} individual which will then interact with all other group representatives during a second phase in an \emph{unbounded confidence} regime. A straightforward interpretation of this two-step decision process is in terms of a delegate system in which like-minded individuals form consensus in groups and then let group representatives negotiate with each other.

A question that arises when considering processes of consensus formation is that of which consensus may be considered the optimal one. Here, we define optimal consensus as the true average of the initial individual opinions, or in other words the collective opinion that cumulatively requires the least change of opinions with respect to the initial state. This can be seen as the simplest democratic optimisation to a problem of finding a common agreement within a population. In this paper, we study in particular under which conditions hierarchical structures -- as well as the social interactions bound to them -- give rise to a collective decision that is closer to the \emph{optimal consensus} than that of a population lacking such structures.

In the following section, we first provide a detailed description of the bounded confidence model as well as of the extension studied in this paper. We then present numerical results for different strengths of the interaction bias, as well as for different sizes and composition of initial groups forming local consensus in the first step of the two-step process. We finally interpret and discuss our results, and comment on their relevance for the field of collective decision making in scenarios in which it is needed to reach a consensus.

\section{Model}
The question of how groups or societies collectively reach decisions has recently gained much attention and valuable insights have been obtained, as illustrated for example by the work of \citet{DyeIoaMorCroCouWatKra2007,DyeJohHelCouKra2009}. Approaches to model such collective phenomena however reach back as far as the early 1990s \citep{GalMos1991}. It is well-known that hierarchical structures play a salient role in such situations and it has been investigated how such structures emerge from the group interactions \citep[see for example the classic Bonabeau model][]{bonabeau1995,bonabeau1996}. Here we do not focus on a study of the mechanisms by which hierarchical structures may emerge, but rather we assume a fixed hierarchical scheme in the decision making process (may it be imposed externally or as a result of social interactions), and study its influence on the eventual consensus decision reached among the individuals of a group.

\subsection{The Bounded Confidence Model}
In this paper we study an extension of the {\em bounded confidence model}. This agent-based model has been proposed independently by Hegselmann and Krause \citep{Krause2000,HegKra2002} as well as Deffuant and Weisbuch \citep{DefNeaAmbWei2000,WeiDefAmbNad2002}. It is suited for studying the dynamics of individual opinions in situations where agents interact to form collective opinions and, in certain cases, consensus (\citet{Lorenz2007} provides a comprehensive review). Let us now introduce the general framework for a continuous bounded confidence model.

We consider a system composed of $N$ individuals, each of which has a {\em continuous} opinion about a topic $x_i(t)$ defined in the interval $[0,1]$, i.e. $x_i(t) \in [0,1]$. Every individual at time $t$ modifies their opinion according to
\begin{equation}
  \frac{d}{dt} x_i(t) = \sum_{j=1}^{N} \kappa( x_j(t)-x_i(t) ) \, \left( x_j(t) - x_i(t) \right).
  \label{eq:xi_dyn}
\end{equation}
Here, the function $\kappa(\delta)$ determines the strength of interaction for two individuals whose opinions are at a distance $\delta$. The model assumes that there is exchange of opinion only if the opinions of both individuals depart from each other {\em at most} a distance $\epsilon$:
\begin{equation}
  \kappa(\delta) =  \zeta \,  H( \epsilon - | \delta |  ),
  \label{eq:interaction}
\end{equation}
where $H$ is the Heaviside function, $H(\cdot) = 1$ if its argument is positive, zero otherwise. The parameter $\zeta$ is the strength of the interaction. It has been shown that the bounded confidence model displays a stationary state with $\sim [ 1/2\epsilon ]$ clusters, if the initial condition is uniform in the unit interval \citep{DefNeaAmbWei2000,BenKraRed2003}.

\subsection{Opinion Bias and Hierarchical Structure}
In public decision making, the individuals may be subject to a bias in the opinion formation process. Such bias may be the result of internal conviction, media influence, or the preservation of their {\em status quo} \citep{GalMos1991}. In this setting, is it possible that the election of representatives may alleviate the bias and make the population reach a consensus closer to the unbiased result? In order to answer this question, we extend the bounded confidence model to include opinion bias and a hierarchical scheme for the creation of representative agents.

There has been work on the hardening of positions when agents stick to their own opinion \citep{friedkin1990,friedkin1999}, and extensions of the bounded confidence model have been proposed to study the influence of heterogeneous confidence thresholds \citep{weisbuch2003,Lorenz2007}, but to our knowledge no studies have been done on the influence of an opinion bias in the bounded confidence model. Whilst the original Deffuant model \citep{DefNeaAmbWei2000} includes a convergence speed $\mu$, or ``cautiousness'' parameter, and some work \citep{laguna2004,assmann2004} showed that the role of this parameter goes beyond a mere time scaling of the convergence, no work has concentrated explicitly on the influence of this convergence speed, especially with regard to a possible asymmetric character for the convergence (which characterises the bias). Likewise, in spite of past empirical studies focusing on the impact of hierarchical structures (compared to egalitarian ones) in decision making processes \citep{edge1984}, to our knowledge there is no analysis linking such considerations to a known opinion dynamics framework.

Our model works as follows: the initial population interacts in a bounded confidence scenario with an interaction threshold $\epsilon_1$. Initially, the individuals opinions are drawn uniformly over the interval $[0,1]$. There is an internal bias in the population that favours one of the two extreme opinions $\{0,1\}$. In order to model this behaviour, the interaction term is given by
\begin{equation}
\kappa(\delta) =
\begin{cases}
  \zeta \,   & \text{if} \, 0 > \delta > \epsilon_1 \\
  (\zeta + \sigma ) & \text{if} \, -\epsilon_1  <  \delta < 0
\end{cases},
\label{eq:biased_kappa}
\end{equation}
where $\sigma$ is the level of bias in the system.
Under these conditions, the system is left to evolve for a time $T_1$, a time constant long enough to allow the system to reach the stationary state.

Then, the clusters of individuals who have reached the consensus internally are replaced by one individual, a \textit{representative}, independently of the group size. The system size at this second stage is equal to the number of groups $K$ formed in the first stage, and each representative has an initial state equal to the (local) consensus reached in the previous round. In this second stage, a new threshold for interaction $\epsilon_2>\epsilon_1$ is selected. Here we always choose $\epsilon_2=1$. This condition is sufficient if one wants to make sure that one single consensus will be reached. This second stage in the dynamics constitutes what can be called {\em unbounded confidence}. Choosing instead a value of $\epsilon_2$ such that $\epsilon_1<\epsilon_2<1$ would be equivalent to considering a hierarchical decision making process including multiple (more than two) steps. As we concentrate here on the scenario where consensus need be reached eventually, the final value of the interaction threshold (here $\epsilon_2$) has to be $1$.
In this second stage, the dynamics of the system is driven by the new threshold for interaction $\epsilon_2$, and the model runs for a time $T_2$ until the final state is reached. It is important to note that the effect of the bias is still present in this level, as the representatives are also individuals, and as such also subject to the same conditions as the initial population. It follows that the dynamics of the representatives is given by
\begin{equation}
  \frac{d}{dt} y_i(t) = \sum_{j=1}^{K} \kappa( y_j(t)-y_i(t) ) \, \left( y_j(t) - y_i(t) \right),
\end{equation}
where the coupling is also given by Eq.~\ref{eq:biased_kappa}.

\begin{figure}[t!]
\begin{center}
\includegraphics{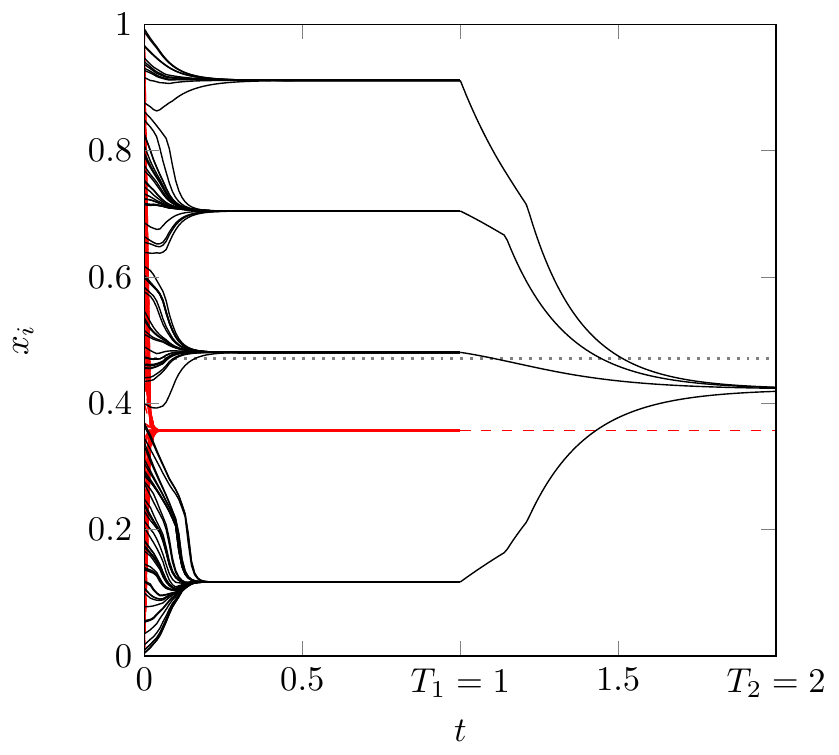}
\end{center}
\caption{Illustration of the model's dynamics. Black lines represent a typical realisation of the hierarchical decision making model, with $N=100$, $\epsilon_1=0.1, \epsilon_2=1$, $\sigma=1$. These dynamics can be compared to the ``best'' possible decision, i.e. the initial average opinion $X_0$ (grey dotted line), and to the output of the non-hierarchical model (classical bounded confidence model, red dashed lines). In each case, both the hierarchical and non-hierarchical realisations were computed with the same initial distribution $x_i(0)$, so as to avoid stochastic bias.}
\label{fig:dynamics}
\end{figure}

\subsection{Measures}
In order to quantify the behaviour of the system, we first compute the initial average opinion of the population,
$$
X_0 = \frac 1 N \sum_{i=1}^{N} x_i(0).
$$
As developed above, $X_0$ is the {\em optimal consensus value}, as it minimises the cumulative opinion deviation of all the agents.
We also measure
the {\em final average opinion}, $Y_2$, computed over the final state of the representatives,
$$
Y_2 = \frac 1 K \sum_{i=1}^{K} y_i(T_2).
$$
Then, a measure for the {\em final error} $E_2(\epsilon_1, \epsilon_2)$ with respect to the initial opinion of the population is given by
$$
 E_2(\epsilon_1, \epsilon_2) = | Y_2 - X_0 |.
$$
In order to be able to assess the benefit of introducing a hierarchical structure (i.e. comparing the two-step with the one-step process), we compute the {\em quality ratio} of the final consensus state
$$
G(\epsilon_1, \epsilon_2) = \frac{  E(\epsilon_1, \epsilon_2) }{ E(\epsilon_2, \epsilon_2) }.
$$
This metric is the ratio between the result achieved by means of the hierarchical process and that which would have been obtained in a fully-unbounded context.

Finally, we compute the total interaction, ${\mathcal H}_1$ and ${\mathcal H}_2$ (for the first and second stages of the process), as a proxy for the amount of opinion exchange. ${\mathcal H}_1$ and ${\mathcal H}_2$ are computed respectively as
$$
{\mathcal H}_1 = \int_0^{T_1} dt\,  \sum_{i=1}^{N} \sum_{j=1}^{N} \kappa( x_i(t) - x_j(t) ),
$$
and
$$
{\mathcal H}_2 = \int_{T_1}^{T_2} dt\,  \sum_{j=1}^{K} \sum_{j=1}^{K} \kappa( y_i(t) - y_j(t) ).
$$
It is worth mentioning that once consensus is reached in each stage, the interaction terms vanish. Hence if $T_1$ and $T_2$ are large enough, the final result does not change, and this measure is well-defined. The total exchange of opinion in the population, for the complete process, is then given simply by ${\mathcal H} = {\mathcal H}_1+{\mathcal H}_2$.

\begin{figure}[t!]
\label{fig:E2}
\begin{center}
\includegraphics{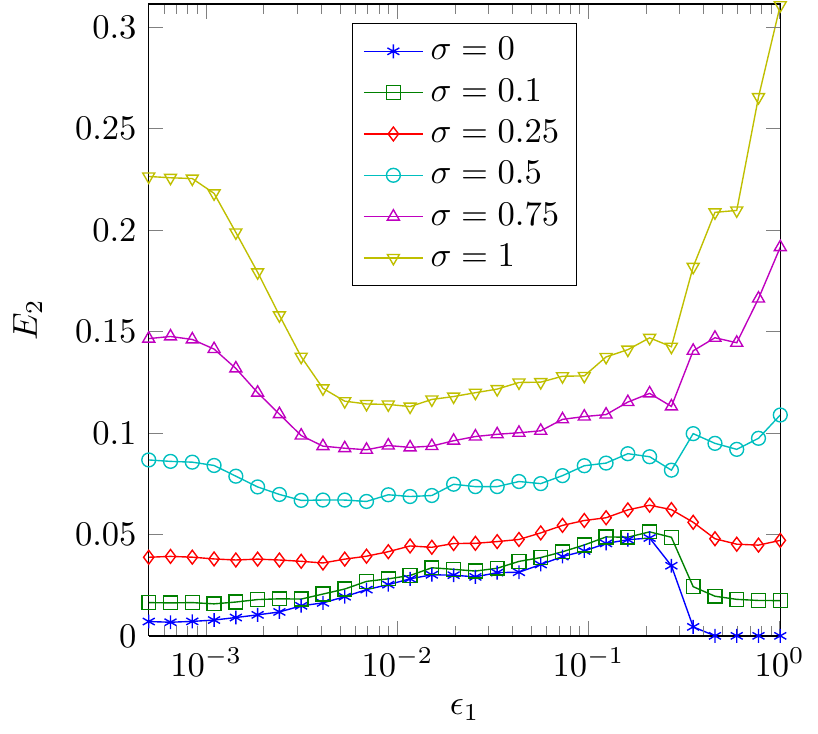}
\end{center}
\caption{Error of the two-step decision model after the final consensus is reached: $E_2(\epsilon_1,\epsilon_2)=|Y_2-X_0|$ (with $\epsilon_2=1$). Lower values of $E_2$ correspond to a higher accuracy of the final decision. For large enough values of $\sigma$, the error is minimised when local hierarchies are formed in the first phase, i.e. for intermediate values of $\epsilon_1$. We used a population of $N=100$ agents. In this figure and the following ones, we computed 1000 realisations of the decision process with different initial distributions and plotted the average value of the metric considered over all realisations.}
\label{fig:error}
\end{figure}

\section{Results}
Figure~\ref{fig:dynamics} illustrates the time evolution of the reputation of 100 agents, all subject to a strength of interaction $\zeta=1$ and an opinion biased by the same factor ($\sigma=1$) towards zero. It can be seen that a two-step decision process may in this case be beneficial, as it produces an eventual consensus that is closer to the optimal consensus $X_0$ than the one produced by the unbounded confidence model, under the same conditions. The final error, i.e. the distance between the optimal consensus and the final decision reached, is presented in Figure~\ref{fig:error}. We see that, logically, stronger values of the bias $\sigma$ produce larger errors. However it is also interesting to note that, for sufficient values of $\sigma$, the error due to opinion bias is mitigated by the formation of non-trivial local hierarchies. We call non-trivial hierarchies those formed by a value of $\epsilon_1\gtrapprox\frac{1}{2N}$ (for lower values of $\epsilon_1$ very few to no groups are formed in the first decision phase) and $\epsilon_1\ll 1$ (if $\epsilon_1\approx 1$, the step-wise model roughly amounts to the unbounded confidence model). Notwithstanding the positive effect of a hierarchical decision process at high values of $\sigma$, we also observe that under weaker opinion bias (e.g., $\sigma=0.1$), the performance of the system is worse when forming local hierarchies ($\frac{1}{2N}\lessapprox\epsilon_1\ll1$) than when using the classical unbounded confidence model, where all agents interact with each other.

\begin{figure}[t!]
\label{fig:G}
\begin{center}
\includegraphics{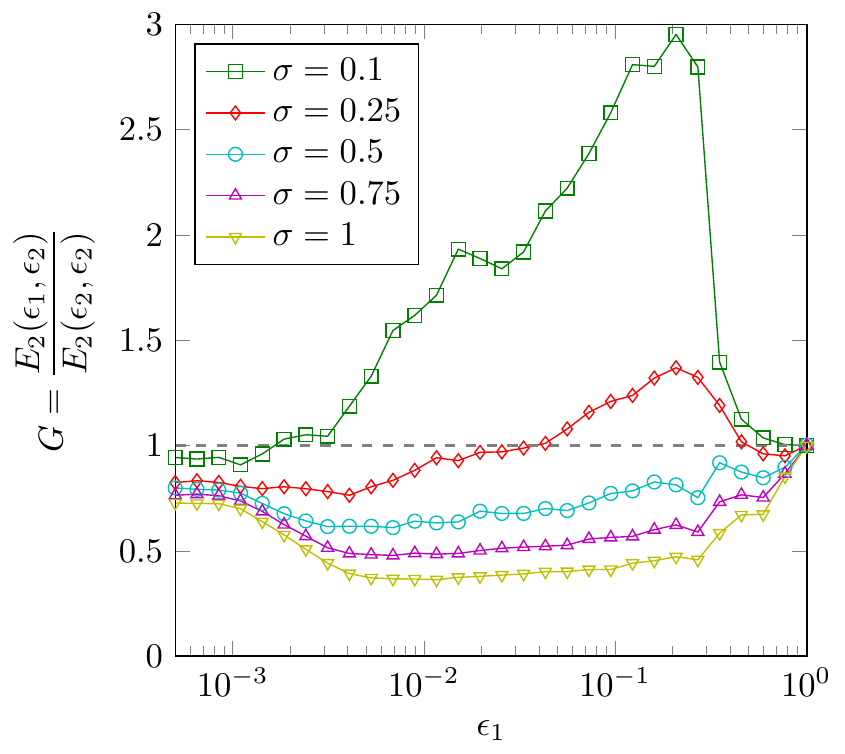}
\end{center}
\caption{Quality ratio of the two-step decision model compared to the unbounded confidence model with equivalent parameters. The dashed line lies at a value of 1, which would amount to an equal performance (in terms of decision accuracy) of the step-wise and the unbounded confidence model. With $\sigma=0$ and $\epsilon_1\to0$, $E_2\to0$ (the unbounded confidence model without opinion bias produces a final consensus that is optimal), and the values of $G$ diverge, which is why $\sigma=0$ is not included here. We used a population of $N=100$ agents.}
\label{fig:quality}
\end{figure}

\begin{figure}
\label{fig:K1}
\begin{center}
\includegraphics{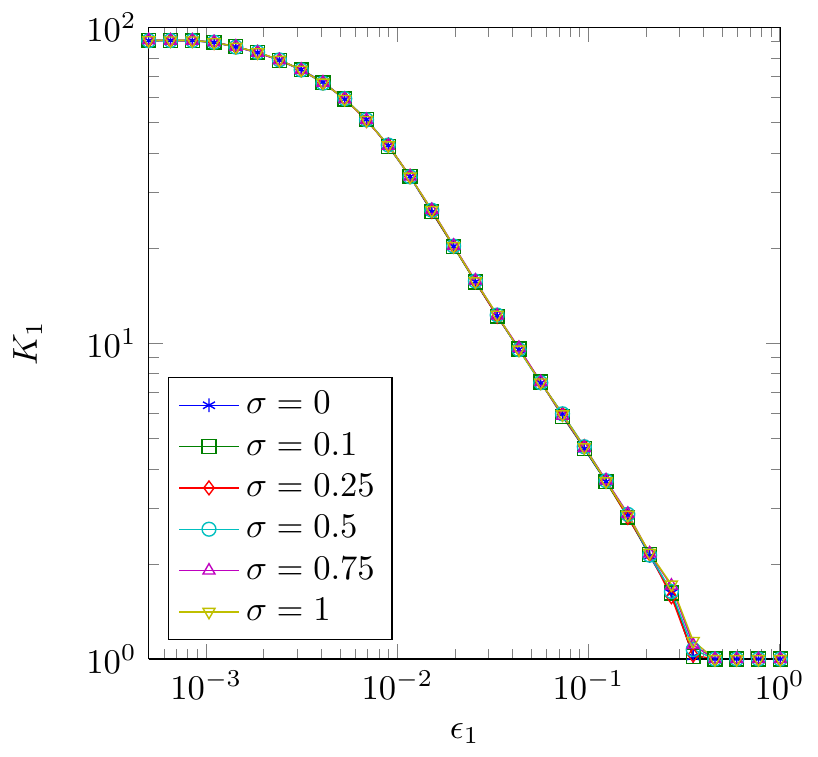}
\end{center}
\caption{Number of clusters formed at $t=T_1$, the end of the first decision phase. We used a population of $N=100$ agents. Because the initial distribution of opinions is subject to stochastic fluctuations, even at low values of $\epsilon_1$ a few local clusters form and $K_1$ never reaches $N$. $\frac{K_1}{N}$ is also the average number of agents that ``elected'' a given representative in the second phase.}
\label{fig:clusters}
\end{figure}

The exact extent of this gain in decision accuracy (or lack thereof) is considered in Figure~\ref{fig:quality}, where we compute the ratio of the final error $E_2$ in the two-step decision model over the error obtained from the unbounded confidence model, for the same parameters. An equal performance of the two models would translate into values of this ratio $G$ around unity. We find however a different picture, with the performance of the two-step decision process varying between consistently worse than the one-step decision process, or unbounded confidence model ($\epsilon_1=\epsilon_2=1$) under low opinion bias, and always better (up to about 2.5 times better for $\sigma=1$ and $\epsilon_1\approx10^{-2}$ ) under stronger opinion bias. This effect is of course only present for non-trivial hierarchies, and the decisional structure created by such a decision making model can be assessed from Figure~\ref{fig:clusters}, where $K_1$ represents the number of local decision clusters formed at the end of the first phase of decision making; it can also be said that $\frac{K_1}{N}$ is the average number of agents whose opinion is aggregated and represented for by each representative in the second phase. In the population of 100 agents we consider, the optimal gain in decision accuracy is found in the range $10^{-2}\leq\epsilon_1\leq10^{-1}$. This amounts to each representative being ``elected'' by a group of 3 to 30 agents at the end of the first phase.

\begin{figure}
\label{fig:H1+H2}
\begin{center}
\includegraphics{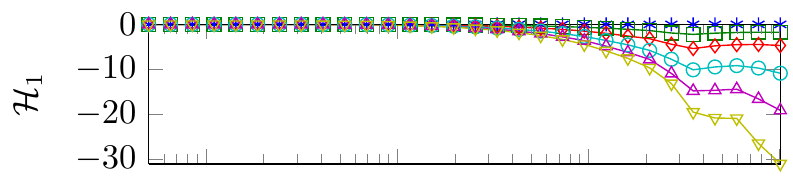}\\
\includegraphics{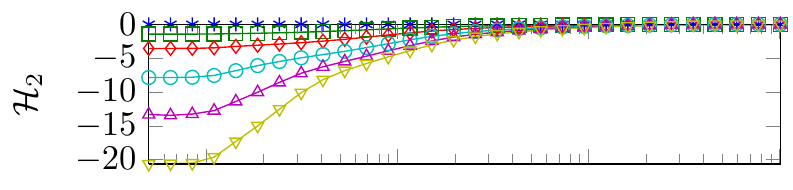}\\
\includegraphics{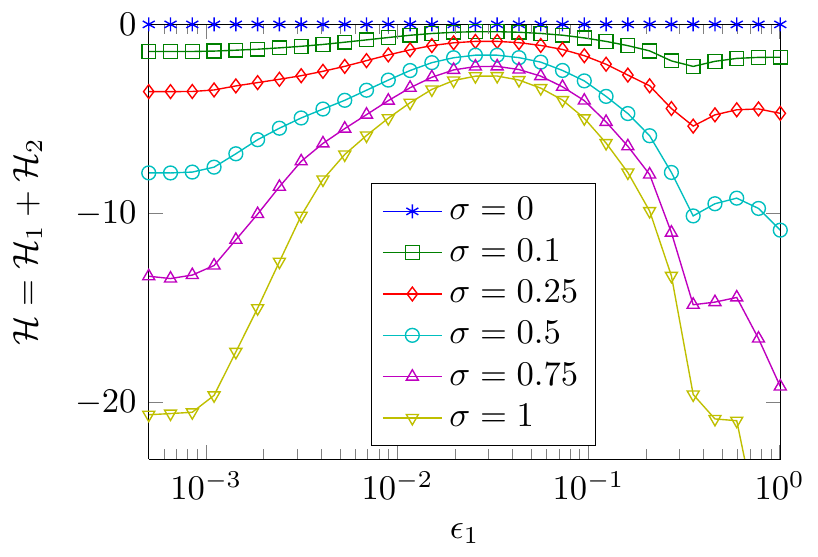}
\end{center}
\caption{Cumulative integration strength in the first decision phase $\mathcal{H}_1$ (top), in the second decision phase $\mathcal{H}_2$ (middle), and the sum of both (bottom panel). We used a population of $N=100$ agents.}
\label{fig:intstrength}
\end{figure}

In the Discussion section, we comment on the importance of the cumulated strength of interaction between agents for the influence of the bias on the eventual consensus reached. Figure~\ref{fig:intstrength} shows both $\mathcal{H}_1$, $\mathcal{H}_2$ (see Measures), as well as their sum. We observe that a stronger bias logically results in a higher cumulated strength of interaction, thereby driving the curves of $\mathcal{H}_1$ and $\mathcal{H}_2$ downwards for increasing $\sigma$ in our example, where the bias drives all opinions toward zero.
We observe that for low values of $\epsilon_1$, very few interactions take place in the first decision phase and most happen in the second phase (if few or no clusters are formed before $T_1$, the second phase will involve most of the convergence toward consensus, hence $\mathcal{H}_1\ll\mathcal{H}_2$. Conversely, at high values of $\epsilon_1$, a global consensus is found even before $T_1$ and we observe $\mathcal{H}_1\gg\mathcal{H}_2$. Of more interest is what happens between those two regimes; as explained in the Measures section, the total exchange of opinion within the population during the decision making process (i.e. the total influence of the opinion bias) is expressed by the sum $\mathcal{H}=\mathcal{H}_1+\mathcal{H}_2$. We observe that this sum finds its minimum (in absolute value) in the region where $\epsilon_1$ allows for the formation of non-trivial hierarchies. This not only supports our insights on the minimisation of interactions for decision accuracy, but also provides interesting insights into the optimal hierarchical decision structure needed to reach an optimal consensus under strong opinion bias.

\vspace{1cm}
\section{Discussion}
The results presented in the previous section show that, at least in the situations and under the assumptions considered in this paper, a two-step consensus formation reduces the impact of an opinion bias on the eventual consensus reached by a group through active interaction, and thus improves the overall quality of distributed decision making processes. In the following, we have a deeper look at why this is the case and what conclusions one can draw from these results.

We first recall that the systemic bias $\sigma$ considered in our model only affects opinion convergence when individuals interact, based on their given distance threshold $\epsilon_1$ or $\epsilon_2$ (depending on which phase of the decision process we are in). As such, the total bias present in the eventual collective decision depends on the number of interactions that can take place based on the confidence interval. Due to the fact that -- in case an interaction actually takes place -- the bias $\sigma$ is multiplied by the difference of opinions (see Equation~\ref{eq:xi_dyn}), the bias present in the eventual decision further depends on the differences in the opinions of interacting agents. In particular, this means that the final bias also depends on the time needed for the opinions to converge.
Based on this, an interpretation of our results is that in situations where a bias coupled to interactions negatively affects the quality of the eventual collective decision, a hierarchical decision process as studied in this article provides an optimal trade-off in terms of the cumulative interaction necessary to reach consensus and the total bias introduced in the process of decision making. This, however, only holds for strong opinion bias, as discussed below.

To underpin this interpretation of our results, we have measured the \emph{cumulative interaction} ${\mathcal H}$ taking place during both steps of the hierarchical consensus formation process. Intuitively, ${\mathcal H}$ integrates the number of interactions as well as the cumulative opinion change subject to the systemic bias over time. In general, since each interaction introduces a bias, a minimisation of ${\mathcal H}$ under the constraint that a single collective decision still emerges should minimise the final deviation from the optimum value.
${\mathcal H}_1$ and ${\mathcal H}_2$ -- which measure the cumulative interactions in step one and two of the hierarchical process, respectively --, as well as  ${\mathcal H}$ -- which sums the interaction in both steps -- are shown in Figure~\ref{fig:intstrength} for different values of the confidence interval $\epsilon_1$.
One observes in the top panel of Figure~\ref{fig:intstrength} that, during the first phase of the two-step process, virtually no interaction takes place for small confidence intervals $\epsilon_1<10^{-1}$.
Figure~\ref{fig:clusters} shows that this results in many small clusters being represented by many representative agents in the second step of the two-step process.
As shown in the middle panel of Figure~\ref{fig:intstrength}, this leads in turn to a large amount of cumulative interactions (in absolute terms) during the second phase of the process, and thus to a large bias in the collective decision.
Only for an intermediate regime ($10^{-2}\leq\epsilon_1\leq10^{-1}$, bottom panel of Figure~\ref{fig:intstrength}) are cumulative interactions (and thus the bias introduced) minimised, thus resulting in a low error. 

A very important result can be seen in Figure~\ref{fig:error}, for small values of the opinion bias $\sigma$. In this setting, the final error $E_2$ for intermediate values of the parameter $\epsilon_1$ is actually larger than in the unbounded confidence context. The reason for this effect is that the formation of the hierarchy causes the system to lose part of its averaging power. This can be seen as a mitigation of the ``wisdom of crowds'' effect \citep{surowiecki2004}: indeed, when $K$ groups are formed in the first stage, each group is composed of roughly $\sim N/K$ individuals. If this number is small, the local average of each group is subject to large fluctuations. A similar argument can be used if the number $K$ is small. Interestingly, a strong enough bias can counterbalance this effect, in the sense that the bias introduced by each interaction outweighs these finite-size fluctuations.

\section{Conclusions}
In this paper we have used a standard model of collective opinion dynamics as a basis to study whether hierarchical consensus formation can improve the accuracy of a distributed decision making process, in which each interaction is subject to a bias, which can be seen as the effect of media on the population, or more generally as a common source of polarised influence on all the individuals.

Our findings show that the strength of the bias may lead to different results with respect to whether hierarchical consensus formation leads to better collective decisions in terms of the final deviation from the true average. For a small bias, the reduction of interactions that is due to the hierarchical organisation may turn out to be detrimental in terms of an increased error. For strong biases, we find that a hierarchical decision structure is always better, with an optimum value for the confidence threshold which is independent of the bias strength.

These results foreshadow several possible extensions of this work; one is to study how these results depend on the population size. For larger systems, it is important to address how many levels the hierarchy should comprise in order to maximise the benefits of a hierarchical decision making structure (multiple-level hierarchies can have non-trivial implications, as underlined for example by \citet{galam2000} in a slightly different context). In this generalised setting, the relation between a level $l$ in the hierarchy and the corresponding $\epsilon_l$ should be discussed. Furthermore, in this work we have considered an exogenous bias. It would be interesting to study the role of an endogenously generated bias, by linking the bias with the current average opinion of the population. In this scenario, there would likely be a self-reinforcing dynamics that amplifies small initial opinion fluctuations in the population. Another possible extension could be to consider a non-uniform response to the bias depending on the location of an individual in the opinion space, related to the asymmetric confidence introduced by \citet{HegKra2002}. This would allow a more realistic study of opinion dynamics when influenced by an extremist minority, which finds direct applications in voting scenarios. To the best of our knowledge, such extensions have not been considered from the point of view of hierarchical decision making structures. Additionally, a certain level of analytical reduction is possible with the bounded confidence model \citep{HegKra2002,Lorenz2007}; further investigation could focus on an analytical treatment of the model presented here in order to obtain general results on the usefulness of electing representatives in the context of consensus formation under opinion bias.

In summary, we have studied the influence of social interactions and hierarchical structures on the quality of group decision making processes. The problem of consensus formation in heterogeneous populations and under diverse conditions is a very topical, which is the focus of active research at the moment \citep{conradt2005,DyeIoaMorCroCouWatKra2007,DyeJohHelCouKra2009,couzin2011}. Whilst in this paper we focused on a very specific scenario and limited the analysis to a single (albeit standard) model, we explored a direction that to our knowledge had remained hitherto uncharted and we expect that future studies on the topic will follow this first step. In general, we think that work along this line of research is crucial for a substantiated understanding of collective decision making.

\bibliographystyle{sg-bibstyle}
\begin{small}
\bibliography{Paper_ECMS2012}
\end{small}

\end{document}